# Data-Driven Distributionally Robust Scheduling of Community Integrated Energy Systems with Uncertain Renewable Generations Considering Integrated Demand Response

Yang Li [a,*], Meng Han [b], Mohammad Shahidehpour [c], Jiazheng Li [d], Chao Long [e]

[a] School of Electrical Engineering, Northeast Electric Power University, Jilin 132012, China

[b] State Grid Zibo Power Supply Company, Zibo 255022, China

[c] ECE Department, Illinois Institute of Technology, Chicago, IL 60616, USA

[d] State Grid Xiamen Power Supply Company, Xiamen 361004, China

[e] School of Water, Energy and Environment, Cranfield University, UK

* Corresponding author. E-mail address: liyang@neepu.edu.cn (Y. Li).

**Abstract:** A community integrated energy system (CIES) is an important carrier of the energy internet and smart city in geographical and functional terms. Its emergence provides a new solution to the problems of energy utilization and environmental pollution. To coordinate the integrated demand response and uncertainty of renewable energy generation (RGs), a data-driven two-stage distributionally robust optimization (DRO) model is constructed. A comprehensive norm consisting of the 1-norm and ∞-norm is used as the uncertainty probability distribution information set, thereby avoiding complex probability density information. To address multiple uncertainties of RGs, a generative adversarial network based on the Wasserstein distance with gradient penalty is proposed to generate RG scenarios, which has wide applicability. To further tap the potential of the demand response, we take into account the ambiguity of human thermal comfort and the thermal inertia of buildings. Thus, an integrated demand response mechanism is developed that effectively promotes the consumption of renewable energy. The proposed method is simulated in an actual CIES in North China. In comparison with traditional stochastic programming and robust optimization, it is verified that the proposed DRO model properly balances the relationship between economical operation and robustness while exhibiting stronger adaptability. Furthermore, our approach outperforms other commonly used DRO methods with better operational economy, lower renewable power curtailment rate, and higher computational efficiency.

## 1 Introduction

In recent years, fossil fuel depletion and environmental pollution have become two major problems that are being faced by the entire human society. The emergence and development of renewable generations (RGs) provide an opportunity to address such problems [1]. A community integrated energy system (CIES) is a typical example and important support for the development of energy internet [2]. A CIES mainly establishes the power system as the basic framework, realizes the stepped utilization of energy by coupling various energy forms such as electricity, gas, and heat, and gives full play to the complementary characteristics of various heterogeneous energy sources [3,4]. However, with the increasing penetration rate of renewable energy in CIESs, its inherent uncertainty will inevitably impact the stable operation of the CIESs and decrease the consumption rate of renewable energy [5]. Therefore, managing the uncertainty of RGs has become a hot research topic in CIES optimization operation. Currently, stochastic programming and robust optimization are two common approaches to dealing with uncertainties in CIESs [6]. In [7], a robust method was adopted to deal with the uncertainty of energy market prices and predict confidence intervals of uncertain parameters by a Gaussian process method. In [8], an IES stochastic optimization model was proposed; this model incorporates network dynamic characteristics and psychological preferences to handle the uncertainty and spatiotemporal correlation between multiple wind farms. Ref. [9] proposed a day-ahead stochastic optimization model for to manage the uncertainty of forecasting distributed generations and loads by using Latin hypercube sampling and the K-means method for scenario generation and clustering, respectively. Unfortunately, in practical applications, the above methods have certain limitations. Stochastic programming usually assumes that random variables must obey a certain deterministic distribution, but often probability distribution models cannot accurately describe the real distribution of actual RGs, and the large scale of discrete scenarios leads to long computation times [10]. Robust optimization utilizes an uncertainty set to describe the fluctuation range of uncertain parameters, but it often leads to conservative optimal solutions as a result of too much emphasis put on the worst scenarios [11]. The emerging distributionally robust optimization (DRO) combines the characteristics of both methods. It does not need to set the type and parameters of the probability distribution; it can establish an uncertain set of probability distributions based on data and target the worst probability distribution to make decisions. The general distributionally robust method usually adopts the probability distribution ambiguity set based on distance [12] and moment information [13], but the resulting NP-hard problem is difficult to solve. In [14], the authors established the probability distribution ambiguity set of RGs based on the Wasserstein distance, and addressed the DRO problem of optimal power flow in transmission and distribution networks. Given that there may be a deviation between the statistical moment of the probability distribution of random variables and the set value, the probability distribution ambiguity set of RGs in [15] was based on the moment information, which was adjusted according to the risk preference of decision makers. The larger the risk parameter value, the more conservative the scheduling scheme.

In recent years, artificial intelligence technology has been gradually applied to optimal scheduling of power systems [16-18]. The traditional method for probability model scenario generation lacks wide applicability in complex practical application environments. However, leveraging on the deep-learning framework, deeply hidden statistical patterns in the data can be mined to generate unsupervised scenarios. In [19], generative adversarial networks (GANs) were used to generate diversified contingency scenarios to train a deep learning classifier. Compared with other types of scenario generation methods, the deep-learning-based generation method provides stronger generalization and data expression abilities. It is also advantageous for unsupervised and autonomous learning and can better reflect the spatiotemporal characteristics of random variables, which can improve the ability of systems to manage future uncertainties [20].

To motivate users to actively participate in system scheduling, the demand response optimizes flexible loads on the demand side, thereby achieving the goal of promoting renewable energy consumption and optimal economical operation [21]. In [22], a load management strategy based on the type of load composition was proposed along with an IES multi-objective operation optimization model including an electricity and heat demand response mechanism. In [23], the authors proposed an optimal operation method considering multiple uncertainties and price-based demand response for CIESs; however, incentive-based demand response was not taken into account.

The traditional demand response does not fully exploit the complementary coupling relationship between different energy sources, and previous studies considering the heat demand response often simply treated the controllable heat load as a function of the proportion of the total heat load or outdoor temperature, and did not exploit the ambiguity of user's thermal comfort [24, 25].

Aiming to address these problems, this paper proposes a data-driven two-stage distributionally robust CIES optimal scheduling model. Compared to existing research, the main innovations and contributions of this study are as follows:

(1) To coordinate the integrated demand response (IDR) and uncertainty of RGs, a data-driven two-stage distributionally robust CIES scheduling model was constructed. The traditional DRO method based on the moment information and the probability density is generally solved by constructing a convex optimization model through the deflected linear decision rule and the duality theorem, which increases the difficulty of model transformation and the corresponding NP hard problem is complex. The DRO method proposed in this paper has strong practical values because it only needs a simple linearization and avoids complex mathematical conversions and calculation processes.

(2) The original GAN is more inclined to guide the generator for generating a single RG power distribution with the highest probability. In order to overcome this shortcoming, a scenario generation method based on Wasserstein GAN with gradient penalty (WGAN-GP) is proposed, where WGAN-GP has a better performance in improving the worst case probability distribution estimation according to the sampled data. It can also reflect the space-time characteristics of random variables better and provide a better adaptability in actual system operations.

(3) To further tap the potential of demand response, we comprehensively analyzed the ambiguity of human thermal comfort and thermal inertia of buildings. As a result, an integrated demand response mechanism was developed that effectively promotes the consumption of renewable energy.

(4) A simulation test was carried out on an actual CIES in North China. In comparison with traditional stochastic programming and robust optimization methods, it was verified that the proposed DRO model better balances the relationship between economical system operation and conservatism, and exhibits better performance and stronger adaptability. Furthermore, comparisons have been made between our proposed solution and other commonly used DRO methods to examine the superiority of our proposed DRO.

## 2 RGs scenario generation and reduction
### 2.1 RGs scenario generation based on WGAN-GP

The generative adversarial network (GAN) is an unsupervised machine learning method which was proposed by Goodfellow et al. in 2014. It consists of generator $G$ and discriminator $D$ [26]. $G$ is a sample generator. When a group of noise vectors $z$ is input, $G$ generates new data samples $G(z)$ by learning the inherent distribution laws and characteristics of real data $x$; $D$ is a two classifier, and its input is the data generated by $G$ and the real data $x$. The purpose of generator $G$ is to generate samples that are as distributed as the real data to maximize the interference with the discriminator. Discriminator $D$ is designed to distinguish between the generated sample $G(z)$ and the real data $x$. This is the dynamic game process of $G$ and $D$, where both can improve their performances continuously through the iterative confrontation training, until $D$ can no longer distinguish between the real data $x$ and the generated sample $G(z)$; then they reach the Nash equilibrium.

The basic structure of a GAN is shown in Fig. 1.

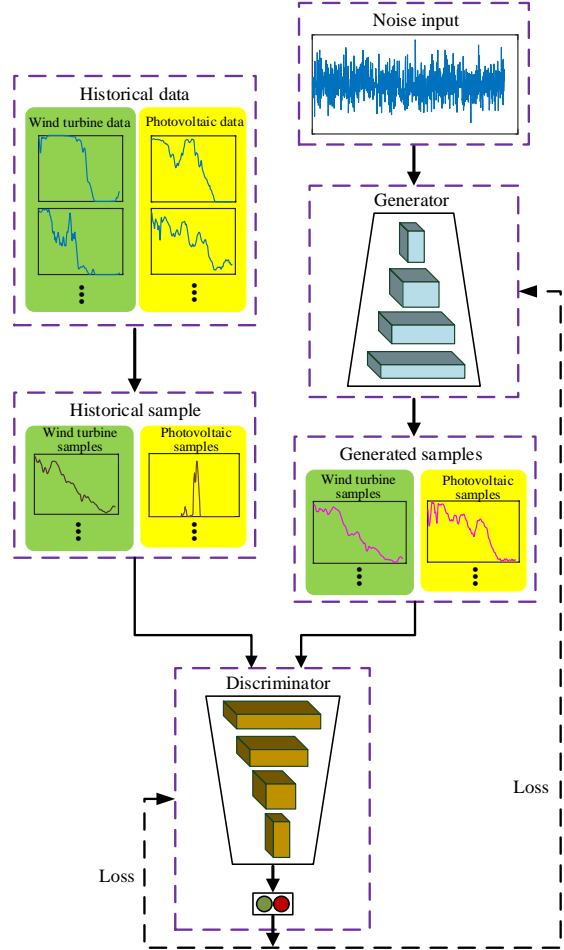

**Fig. 1. Basic structure of a GAN**

For simplicity, assuming that the real historical data of RGs in a CIES is $x$, and its probability distribution is $P_d$. There exists a set of random noise vectors $z$ whose probability distribution satisfies $P_z$. A GAN can establish a mapping relationship between $P_d$ and $P_z$, and the generated samples that satisfy the distribution relationship of the real data by training the generator and discriminator are obtained with probability distribution $P_g$. The training process is completed by two deep neural networks, namely generator $G(z;\theta_G)$ and discriminator $D(x;\theta_D)$, where $\theta_G$ and $\theta_D$ are the weights of the corresponding neural networks. Therefore, the loss functions for both networks can be defined as follows:

$$L_G = E_{z \sim P_z}[-D(G(z))] \tag{1}$$

$$L_D = E_{z \sim P_z}[D(G(z))] - E_{x \sim P_d}[D(x)] \tag{2}$$

where $L_G$ and $L_D$ represent the loss functions of the generator and discriminator, respectively; $E(\cdot)$ is the expectation function; $G(\cdot)$ is the generator function; and $D(\cdot)$ is the discriminator function.

The smaller the value of $L_G$, the closer the samples generated by the generator to the actual data, and the stronger the ability of the discriminator to distinguish the authenticity of the data. The objective function of the game process is expressed as follows:

$$\min_G \max_D V(G,D) = E_{x \sim P_d}[D(x)] - E_{z \sim P_z}[D(G(z))] \tag{3}$$

WGAN uses the Wasserstein distance instead of the Jensen-Shannon distance to measure the distance between real data and generated samples based on the original GAN. This effectively solves the problems of gradient disappearance and training instability in GAN [27]. WGAN-GP solves the problem that WGAN cannot constrain the discriminator within the 1-Lopschitz constraint by penalizing the gradient norm of the WGAN discriminator. The objective function of the overall training of a WGAN-GP is expressed as follows:

$$\min_G \max_D V(G,D) = E_{x \sim P_d}[D(x)] - \\ E_{z \sim P_z}[D(G(z))] + \rho E_{\hat{x} \sim P_{\hat{x}}}[\|\nabla \hat{x} D(\hat{x})\|_2 - 1]^2 \tag{4}$$

where $\|\cdot\|_2$ represents the 2-norm, $\rho$ is the penalty coefficient, and $\hat{x} = \varepsilon x + (1 - G(z))$, where $\varepsilon$ is a random number that satisfies the uniform distribution on the interval [0,1].

**2.2 RGs scenario reduction based on K-means++**

The K-means++ clustering algorithm is widely used in the reduction of RG scenarios in power systems because of its simplicity and efficiency and its better performance than the original K-means algorithm in selecting initial cluster centers [28]. In this study, the K-means++ algorithm was used to cluster a large number of RG scenarios generated by a WGAN-GP. Given that the number of clusters has a great influence on the clustering effect, scheduling model construction, and computational efficiency, it is necessary to determine the optimal number of clusters.

To this end, typical internal validity indexes such as the Davies Bouldin index (DBI) and silhouette coefficient (SC) are introduced for quantitative analysis, both of which are widely used in the study of multi-wind farm correlation, classification of load curves, etc. [29].

Assuming that the $k$ samples in the dataset are clustered into three categories, DBI is expressed as follows:

$$DBI(i) = \frac{1}{N}\sum_{i=1}^{N}\max\left[\frac{S_i + S_j}{d(C_i, C_j)}\right] \tag{5}$$

where $C_i$ is the center of the $i$-th cluster; $S_i$ is the average distance from all samples in cluster $i$ to cluster center $C_i$, also known as intra-cluster distance; and $d(C_i, C_j)$ is the distance between cluster centers $C_i$ and $C_j$, also known as inter-cluster distance.

SC is expressed as follows:

$$s(k) = \frac{b(k) - a(k)}{\max\{a(k), b(k)\}} \tag{6}$$

where $s(k)$ is the silhouette coefficient of the $k$-th sample, $b(k)$ is the average distance between the $k$-th sample and other samples outside the cluster, and $a(k)$ is the average distance between the $k$-th sample and all samples in the cluster.

**3 Physical model of CIES**

## 3.1 Overall structure of CIES

The CIES studied in this paper consists of two energy subsystems. Electric energy is supplied by the micro-turbine generator (MTG), WTs, photovoltaic (PV), and the main grid, while thermal energy comes from the MTG and electric boilers (EBs). Electricity and heat loads are residential electricity load and building heat load, respectively. To ensure the safe, flexible, and economical operation of the system, buffer equipment such as an energy storage system (ESS) and a heat storage device (HSD) are also configured. The overall structure of the CIES is shown in Fig. 2.

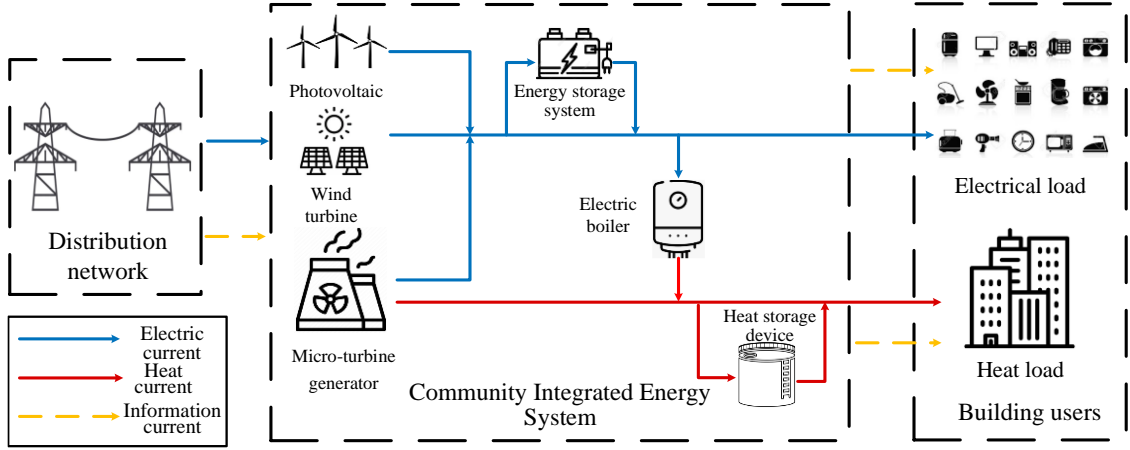

Fig. 2. Schematic diagram of the proposed CIES

## 3.2 Flexible load model
### 3.2.1 Electric load model

According to the response characteristics of the electric load, it is categorized into time-shiftable load (TSL) and electrical interruptible load (EIL).

(1) Time-shiftable load

Under the condition that the total power demand is guaranteed to be fulfilled, the TSL can flexibly adjust the power consumption periods through price guidance to achieve peak shaving and valley filling. It can be described as follows:

$$P_{t,\min}^{TSL} \leq P_t^{TSL} \leq P_{t,\max}^{TSL} \tag{7}$$

$$\sum_{i=1}^{T} P_t^{TSL} = 0 \tag{8}$$

where $P_t^{TSL}$ is the TSL in period $t$, and $P_{t,\max}^{TSL}$ and $P_{t,\min}^{TSL}$ represent the upper and lower limits of the TSL that can be adjusted in period $t$, respectively.

(2) Electrical interruptible load

During the periods of high electricity demand, users would stop power consumption according to the power agreement to ease the power supply pressure during these periods and reduce their own energy consumption cost. The relevant constraints for interruptible loads can be described as follows:

$$0 \leq P_t^{EIL} \leq P_{t,\max}^{EIL} \tag{9}$$

where $P_t^{EIL}$ and $P_{t,\max}^{EIL}$ are the amount of electrical load interrupted and the maximum allowable interruption load in period $t$, respectively. In this study, $P_{t,\max}^{EIL}$ is assumed to be 10% of the electrical load demand in each period. Accordingly, the following expression can be derived:

$$P_{LD,t} = P_{LD,t}^0 + P_t^{TSE} - P_t^{IE} \tag{10}$$

where $P_{LD,t}^0$ and $P_{LD,t}$ represent the electrical load before and after implementing demand response programs.

**3.2.2 Thermal flexible load**

(1) Predicted mean vote

To quantify the user's acceptable thermal comfort range more accurately, a predicted mean vote (PMV) is introduced as follows:

$$PMV = 2.43 - \frac{3.76(T_s - T_{in}(t))}{Q(I_{cl} + 0.1)} \tag{11}$$

where $Q$ is the metabolic rate of the human body; $I_{cl}$ is the thermal resistance of clothing; $T_{in}(t)$ is the room temperature; and $T_s$ is the average temperature when the human skin feels comfortable.

The human body is active during the day and has high requirements for thermal comfort. However, when users are sleeping at night, their comfort requirements can be appropriately relaxed. Therefore, different PMV values are set at different periods [30]; in this study, the indoor temperature values for the whole day were set as shown in Fig. 3. The range of PMV variation is expressed as follows:

$$\begin{cases} |PMV| \leq 0.9, & t \in [1:00-7:00] \cup [20:00-24:00] \\ |PMV| \leq 0.5, & t \in [8:00-19:00] \end{cases} \tag{12}$$

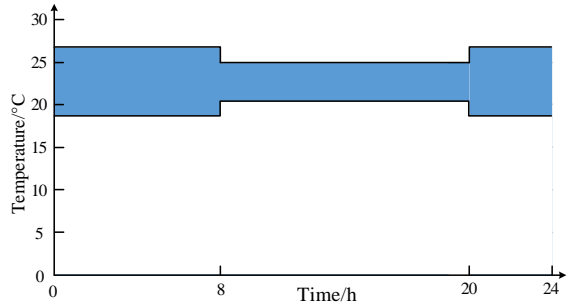

**Fig. 3. The range of indoor temperature variation**

(2) Thermal inertia of buildings

The transient heat balance equation of a building is used to describe the effect of heat change exerted by the heating system on the building temperature; this yields the relationship between the building heat and temperature. The transient heat balance equation for a building can be described through the following first-order differential equation [31]:

$$\frac{dT_t^{in}}{dt} = \frac{H_t^L - (T_t^{in} - T_t^{out}) \cdot K \cdot F}{c_{air} \cdot \rho_{air} \cdot V} \tag{13}$$

where $T_t^{in}$ and $T_t^{out}$ are the indoor and outdoor temperatures in period $t$, respectively; $H_t^L$ is the heating power in period $t$; $F$ and $V$ are the surface area and volume of the building, respectively; $K$ is the comprehensive heat transfer coefficient of the building; and $\rho_{air}$ and $C_{air}$ are the density and specific heat capacity of the indoor air, respectively.

Based on the reasonable assumption that the outdoor temperature remains unchanged for a short time, Eq. (13) can be linearized to obtain the following expression:

$$T_t^{in} = (-\frac{K \cdot F}{c_{air} \cdot \rho_{air} \cdot V} \cdot \Delta t) \cdot (T_{t-1}^{in} - T_t^{out} - H_t^L \cdot \frac{1}{K \cdot F}) + (T_t^{out} + H_t^L \cdot \frac{1}{K \cdot F}) \quad (14)$$

The building heating load is obtained as follows:

$$H_t^L = \frac{[T_t^{in} - T_t^{out}] + \frac{K \cdot F}{c_{air} \cdot \rho_{air} \cdot V} \cdot \Delta t \cdot [T_{t-1}^{in} - T_t^{out}]}{\frac{1}{K \cdot F} + \frac{1}{c_{air} \cdot \rho_{air} \cdot V} \cdot \Delta t} \quad (15)$$

(3) Interruptible heat load

The PMV is used to measure the ambiguity of human body's perception concerning indoor temperature changes, and the range of interruptible heat load can be obtained through the transient heat balance equation. Therefore, the heat supply reduction load can be described as follows:

$$0 \leq H_t^{IE} \leq H_{t,\max}^{IE} \quad (16)$$

$$H_{LD,t} = H_{LD,t}^0 - H_t^{IE} \quad (17)$$

where $H_t^{IE}$ and $H_{t,\max}^{IE}$ are the interrupted heating power and its maximum value in period $t$, respectively; and $H_{LD,t}^0$ and $H_{LD,t}$ represent the heat load demand before and after the reduction, respectively.

## 4 Two-stage DRO model

### 4.1 Two-stage objective function
#### 4.1.1 First-stage objective function

The objective function of the first stage considers the gas turbine startup and shutdown costs:

$$\min \sum_{t=1}^{T} \sum_{i=1}^{N_G} \left( S_{K_i} y_{i,t} + S_{T_i} z_{i,t} \right) \quad (18)$$

where $T$ is the total number of periods in a scheduling cycle, which is set to 24 h; $N_G$ is the number of MTGs; $S_{K_i}$ and $S_{T_i}$ are the startup and shutdown costs of the MTG, respectively; $y_{i,t}$ and $z_{i,t}$ are respectively the MTG startup and shutdown variables. If the values of $y_{i,t}$ and $z_{i,t}$ are 1, it means that the micro-gas turbine is on/off; otherwise, they are 0.

#### 4.1.2 Second-stage objective function

The objective function of the second stage includes the MTG operation cost, main network power purchase cost, ESS operation cost, HSD operation cost, wind and solar curtailment cost, $CO_2$ emission penalty cost, and demand response cost. This objective function is described as follows:

$$\min \left( C_{MTG} + C_{grid}^{buy} + C_{ESS} + C_{HSD} + C_{Loss} + C_{CO_2} - C_{grid}^{sell} + C_{IDR} \right)$$

$$C_{MTG} = \sum_{t=1}^{T} \sum_{i=1}^{N_G} a_i P_{EL,i,t}^{MTG} + b_i u_{i,t}^{MTG}$$

$$C_{grid}^{buy} = \sum_{t=1}^{T} c_{grid}^{buy} P_{grid,t}^{buy}$$

$$C_{ESS} = \sum_{t=1}^{T} \left[ c_{ESS} (P_{CH,t}^{ESS} + P_{DC,t}^{ESS}) \right]$$

$$C_{HSD} = \sum_{t=1}^{T}\left[ c_{HSD}(P_{CH,t}^{HSD} + P_{DC,t}^{HSD}) \right] \tag{19}$$

$$C_{Loss} = \sum_{t=1}^{T} c_{Loss}\left[ \left(\hat{P}_t^{PV} - P_t^{PV}\right) + \left(\hat{P}_t^{WT} - P_t^{WT}\right) \right]$$

$$C_{CO_2} = c_{co_2} \sum_{t=1}^{T}\left( k_{MTG} \sum_{i=1}^{N_G} P_{EL,i,t}^{MTG} + k_{grid} P_{grid,t}^{buy} \right)$$

$$C_{grid}^{sell} = \sum_{t=1}^{T} c_{grid}^{sell} P_{grid,t}^{sell}$$

$$C_{IDR} = \omega_{TSE} P_t^{TSE} + \omega_{EIE} P_t^{IE} + \omega_{HIE} H_t^{IE}$$

where $C_{MTG}$ is the MTG operating cost; $c_{grid}^{buy}$ is the power purchase cost from the main grid; $C_{ESS}$ and $C_{HSD}$ are the operating costs of the ESS and HSD, respectively; $C_{Loss}$ is the cost of curtailing wind and solar power; $C_{CO_2}$ is the $CO_2$ emission penalty cost; $C_{grid}^{sell}$ is the grid electricity sales profit; $C_{IDR}$ is the cost of the demand response; $a_i$ and $b_i$ represent the primary term and constant term coefficients of the MTG operation cost, respectively; $c_{grid}^{buy}$ and $c_{grid}^{sell}$ are the unit prices of electricity purchased and sold by the CIES from and to the grid, respectively; $c_{ESS}$ and $c_{HSD}$ are the unit prices of ESS and HSD charging and discharging operation, respectively; $c_{Loss}$ is the unit penalty cost for curtailing WT and PV; $c_{CO_2}$ is the $CO_2$ emission unit penalty price; $\omega_{TSE}$ is the TSL compensation price; $\omega_{EIE}$ and $\omega_{HIE}$ are the electric and heating compensation prices for interruption loads, respectively; $P_{EL,i,t}^{MTG}$ is the output power of the MTG in period $t$; $\xi_{i,t}$ is the MTG running status flag: if $\xi_{i,t}=1$ means the MTG is running, otherwise $\xi_{i,t}=0$ means the MTG is not running; $P_{grid,t}^{buy}$ and $P_{grid,t}^{sell}$ are the powers purchased and sold by the CIES from and to the grid in period $t$, respectively; $P_{CH,t}^{ESS}$ and $P_{DC,t}^{ESS}$ are the charging and discharging powers of the ESS, respectively; $P_{CH,t}^{HSD}$ and $P_{DC,t}^{HSD}$ are the heat storage and discharge powers of the HSD; $\hat{P}_t^{PV}$ and $P_t^{PV}$ are respectively the predicted output and actual output values of the PV in period $t$; $\hat{P}_t^{WT}$ and $P_t^{WT}$ are the predicted output and actual output values of the WT in period $t$, respectively; and $k_{MTG}$ and $k_{grid}$ are the $CO_2$ emissions per unit electricity produced by the MTG and the grid, respectively.

### 4.2 Constraints

#### 4.2.1 MTG operating constraints
(1) MTG output constraint

The MTG output follows the following constraint [32]:

$$0 \leq P_{EL,i,t}^{MTG} \leq \xi_{i,t} P_{MTG,EL}^{max}, \forall i \in \Omega^{MTG} \tag{20}$$

$$y_{i,t} + z_{i,t} \leq 1$$

where $P_{MTG,EL}^{max}$ is the maximum value of the power supplied by the MTG in period $t$ and $\Omega^{MTG}$ is the set of all MTGs.

(2) MTG ramping constraints

The MTG ramping constraints are given as follows:

$$\xi_{i,t} P_{EL,i,t+1}^{MTG} - \xi_{i,t+1} P_{EL,i,t}^{MTG} \leq \Delta P_{U,max}^{MTG}, \forall i \in \Omega^{MTG}$$

$$\xi_{i,t} P_{EL,i,t-1}^{MTG} - \xi_{i,t+1} P_{EL,i,t}^{MTG} \leq \Delta P_{D,max}^{MTG}, \forall i \in \Omega^{MTG} \tag{21}$$

where $P_{i,t}^{MTG}$ and $P_{i,t+1}^{MTG}$ are the output power of the MTG in period $t$ and the next period, respectively; and $\Delta P_U^{\max}$ and $\Delta P_D^{\max}$ are the maximum allowable rates of the MTG climbing and landslide rates, respectively.

(3) MTG thermoelectric ratio constraints

Engineering practice shows that MTG efficiency is only approximately 30% when it is purely used as a generator, but in the cogeneration mode, the efficiency can reach 75%. Its constraints are defined as follows:

$$P_{HL,i,t}^{MTG} = \eta_{HE} P_{EL,i,t}^{MTG} \tag{22}$$

$$0 \leq P_{HL,i,t}^{MTG} \leq P_{MTG,HL}^{\max}, \ \forall i \in \Omega^{MTG} \tag{23}$$

where $P_{EL,i,t}^{MTG}$ and $P_{HL,i,t}^{MTG}$ are the electrical power output by the MTG and the corresponding heat power, respectively; $P_{MTG,HL}^{\max}$ is the maximum value of the MTG heating power; and $\eta_{HE}$ is the MTG thermoelectric ratio. $\eta_{HE}$ is set as 1.2 in this study, which means that 1 kW of electrical power generated by the MTG will generate 1.2 kW of thermal power.

**4.2.2 ESS operational constraints**

(1) ESS output constraints

The ESS outputs should obey the following constraints [33]:

$$\begin{aligned} 0 \leq P_{CH,t}^{ESS} &\leq B_t^{ch} P_{CH,\max}^{ESS} \\ 0 \leq P_{DC,t}^{ESS} &\leq B_t^{dc} P_{DC,\max}^{ESS} \end{aligned} \tag{24}$$

$$B_t^{ch} + B_t^{dc} \leq 1 \tag{25}$$

where $P_{CH,t}^{ESS}$ and $P_{CH,\max}^{ESS}$ are the charging power of the ESS and its maximum chargeable power, respectively; $P_{DC,t}^{ESS}$ and $P_{DC,\max}^{ESS}$ are the discharge power of the ESS and its maximum dischargeable power, respectively; $B_t^{ch}$ and $B_t^{dc}$ are Boolean variables that represent the running state of the ESS. When the values of $B_t^{ch}$ and $B_t^{dc}$ are 1, the ESS is in a state of charge and discharge, respectively; when their values are 0, the ESS is neither charged nor discharged.

(2) Energy storage capacity constraints

To prolong the lifetime of the ESS and prevent its overcharging and discharging, the following constraints need to be satisfied:

$$C_{\min}^{ESS} \leq C_0^{ESS} + \eta_{ch} \sum_{t=1}^{T} \left( P_{CH,t}^{ESS} \times \Delta t \right) - \frac{1}{\eta_{dc}} \sum_{t=1}^{T} \left( P_{DC,t}^{ESS} \times \Delta t \right) \leq C_{\max}^{ESS} \tag{26}$$

where $C_{\min}^{ESS}$ is the minimum capacity of the ESS; $\eta_{ch}$ and $\eta_{dc}$ are the charging and discharging efficiency of the ESS; and $C_{\min}^{ESS}$ and $C_{\max}^{ESS}$ are the minimum and maximum remaining capacities allowed by the ESS in a scheduling cycle, respectively.

(3) ESS remaining capacity constraints

To ensure a fixed remaining capacity of the ESS after each scheduling period for the scheduling scheme of each period to start from the same state, the following constraints should be satisfied [34,35]:

$$C_0^{ESS} = C_{T_{end}}^{ESS} \tag{27}$$

where $C_0^{ESS}$ is the initial capacity of the ESS, and $C_{T_{end}}^{ESS}$ is the ESS capacity at the end of a scheduling period.

**4.2.3 Power balance constraints**

To maintain the balance of the CIES power supply and demand, the constraints are as follows:

$$P_{grid,t}^{buy} + P_t^{WT} + P_t^{PV} + P_{DC,t}^{ESS} + P_{EL,i,t}^{MTG} = P_{LD,t} + P_{CH,t}^{ESS} + P_{grid,t}^{sell} + P_{EL,t}^{EB}, \quad \forall t \tag{28}$$

where $P_{EL,i,t}^{MTG}$ is the power consumption of the EB.

**4.2.4 Main-grid interactive power constraints**

When the output of the power supply equipment in the CIES is insufficient, the CIES must purchase electricity from the main grid to meet user's needs; conversely, the CIES can sell excess electricity to the grid for profit. The interactive power between the main grid and CIES must satisfy the following constraints:

$$0 \leq P_{grid,t}^{sell} \leq P_{grid,\max}^{sell} \tag{29}$$

$$0 \leq P_{grid,t}^{buy} \leq P_{grid,\max}^{buy} \tag{30}$$

where $P_{grid,\max}^{sell}$ and $P_{grid,\max}^{buy}$ are the upper limits of the interactive powers between the main grid and CIES, respectively.

**4.2.5 Electric-heat power balance constraints**

The heat demand of the CIES is supplied by MTGs and EBs, which is formulated by

$$H_{LD,t} = P_{HL,i,t}^{MTG} + P_{HL,t}^{EB} - P_{CH,t}^{HSD} + P_{DC,t}^{HSD} \tag{31}$$

where $P_t^{HL}$ is the total heat demand of the CIES, and $P_{HL,t}^{EB}$ is the heating power of the EB.

**4.2.6 EB constraints**

The operation of the EB must meet the following constraints:

$$P_{HL,t}^{EB} = \eta_{EB} P_{EL,t}^{EB} \tag{32}$$

$$0 \leq P_{HL,t}^{EB} \leq P_{HL}^{EB} \tag{33}$$

where $P_{HL}^{EB}$ is the rated value of the heating power of the EB and $\eta_{EB}$ is the performance coefficient of the EB, which represents the ratio of heating power to power consumption.

**4.2.7 HSD operating constraints**

(1) HSD output constraints

The HSD output constraints are given as follows:

$$0 \leq P_{CH,t}^{HSD} \leq \beta_t^{ch} P_{CH,\max}^{HSD}$$
$$0 \leq P_{DC,t}^{HSD} \leq \beta_t^{dc} P_{DC,\max}^{HSD} \tag{34}$$

$$\beta_t^{ch} + \beta_t^{dc} \leq 1 \tag{35}$$

where $P_{CH,t}^{HSD}$ and $P_{CH,\max}^{HSD}$ are the HSD heat storage power and its maximum rechargeable power, respectively; $P_{DC,t}^{HSD}$ and $P_{DC,\max}^{HSD}$ are the HSD heat release power and its maximum dischargeable power, respectively; $\beta_t^{ch}$ and $\beta_t^{dc}$ are Boolean variables that represent the HSD operating state. When the values of $\beta_t^{ch}$ and $\beta_t^{dc}$ are 1, the HSD is in the state of storing and releasing heat, respectively; when their values are 0, the HSD neither stores nor releases heat.

(2) HSD capacity constraints

Similar to the ESS, the HSD capacity needs to satisfy the following constraints:

$$C_{t+1}^{HSD} = C_t^{HSD} + (\eta_{ch}^{HSD} P_{CH,t}^{HSD} - P_{DC,t}^{HSD} / \eta_{dc}^{HSD}) \Delta t, \forall t \tag{36}$$

$$C_{\min}^{HSD} \leq C_t^{HSD} \leq C_{\max}^{HSD} \tag{37}$$

where $C_t^{HSD}$ is the ESS capacity; $\eta_{ch}^{HSD}$ and $\eta_{dc}^{HSD}$ are the ESS charge and discharge efficiencies, respectively; $C_{\min}^{HSD}$ and $C_{\max}^{HSD}$ are the HSD allowable minimum and maximum remaining capacities, respectively.

**4.3 Data-driven distributed robust framework**

The first stage consists in formulating the plans for the MTG startup and shutdown, ESS charge and discharge, and HSD heat storage and release. The second stage consists in formulating a corresponding dispatch plan after the RGs uncertainty is revealed, including the MTG and EB outputs, purchase of electricity from the main grid, and sale of electricity. The discrete decision variables that determine the MTG startup-shutdown sign, ESS charge and discharge state variables, and HSD heat storage and release state variables constitute the first-stage variables. Continuous variables such as the MTG output, actual output of WTs and PV, power purchased from the main grid, and output of EBs constitute the second-stage variables. The above decision variables are expressed as follows:

$$U = \left[ y_{i,t}, z_{i,t}, B_t^{ch}, B_t^{dc}, \beta_t^{ch}, \beta_t^{dc} \right]$$
$$V = \begin{bmatrix} P_{EL,i,t}^{MTG}, P_{EL,t}^{EB}, P_{grid,t}^{buy}, P_{grid,t}^{sell}, P_t^{WT}, P_t^{PV}, \\ P_{CH,t}^{ESS}, P_{DC,t}^{ESS}, P_{CH,t}^{HSD}, P_{DC,t}^{HSD}, P_t^{TSE}, P_t^{IE}, H_t^{IE} \end{bmatrix} \tag{38}$$

where $U$ and $V$ represent the sets of decision variables in the first and second stages, respectively.

The two-stage distributionally robust optimization model defined above and described by Eqs. (18)-(37) can be summarized as follows:

$$\min_{u \in U} Au + \max_{p_s \in \Omega^p} \sum_{s=1}^{N_s} p_s \min_{v_s \in V} Bv_s \tag{39}$$

$$s.t. \quad \begin{cases} Cu = c \\ Du \leq d \end{cases} \tag{40}$$

$$\begin{cases} Ev_s = e \\ Fv_s \leq f \end{cases} \forall s \tag{41}$$

$$Gu + Hv_s \leq g \quad \forall s \tag{42}$$

where $p_s$ is the probability of scenario $s$ after clustering; $\Omega^p$ is the set interval of the scenario probability distribution, that is, the confidence interval of the comprehensive norm; $N_s$ is the total number of scenarios after clustering; $A \sim H$ and $c \sim g$ are constant coefficient matrices.

Eq. (40) expresses the equality and inequality constraints of the first-stage decision variables: Eqs. (25) and (35) describe the ESS/HSD operating state constraints; Eq. (41) expresses the equality and inequality constraints of the decision variables in the second stage, specifically those corresponding to Eqs. (20)-(23), (27), (29)-(33), etc., which include constraints of the MTG, ESS capacity, MTG thermoelectric ratio, EB, grid power purchase, and electric-thermal power balance; Eq. (42) expresses the coupling constraint of the decision variables in the first and second stages, which corresponds to the power supply and demand balance constraints described by Eq. (28), and the ESS/HSD charge-discharge/storage-discharge constraints described by Eqs. (24) and (34).

The scenario probability distribution obtained by scenario clustering has a certain error. To make the probability distribution closer to the real data, the uncertainty confidence set presented in this section is centered on the initial probability distribution, and the scenario probability distribution is restricted by the comprehensive norm to obtain the worst probability distribution of each discrete scenario.

The composite-norm probability confidence interval consisting of the 1-norm the ∞-norm is expressed as follows [36]:

$$\Pr\left\{\sum_{s=1}^{N_s}\left|p_s-p_{s0}\right|\leq\theta_1\right\}\geq 1-2N_s\mathrm{e}^{-\frac{2M\theta_1}{N_s}} \tag{43}$$

$$\Pr\left\{\max_{1\leq s\leq N_s}\left|p_s-p_{s0}\right|\leq\theta_\infty\right\}\geq 1-2N_s\mathrm{e}^{-2M\theta_\infty} \tag{44}$$

where $\Pr\{\cdot\}$ is the probability operator; $M$ is the number of historical data; $p_{s0}$ is the initial probability value of the discrete scenario $s$; $\theta_1$ and $\theta_\infty$ are the probability allowable deviation limits under the corresponding constraints of the 1-norm and ∞-norm, respectively.

For convenience of representation, the right sides in Eqs. (43) and (44) are set equal to $\alpha_1$ and $\alpha_\infty$, respectively; here, $\alpha_1$ and $\alpha_\infty$ represent the uncertainty confidences that the probability distribution needs to satisfy. Accordingly, the transformation becomes as follows:

$$\theta_1=\frac{N_s}{2M}\ln\frac{2N_s}{1-\alpha_1} \tag{45}$$

$$\theta_\infty=\frac{1}{2M}\ln\frac{2N_s}{1-\alpha_\infty} \tag{46}$$

From Eqs. (43)-(46), it can be concluded that the confidence set satisfied by the probability distribution is expressed as follows:

$$\Omega^p=\left\{\{p_s\}\left|\begin{array}{l}p_s\geq 0, s=1,\ldots,N_s\\ \sum_{s=1}^{N_s}p_s=1\\ \sum_{s=1}^{N_s}\left|p_s-p_{s0}\right|\leq\theta_1\\ \max_{1\leq s\leq N_s}\left|p_s-p_{s0}\right|\leq\theta_\infty\end{array}\right.\right\} \tag{47}$$

Note that from Eq. (47), the comprehensive-norm confidence interval contains absolute value constraints, so it needs to be linearized and equivalently transformed by introducing 0-1 auxiliary variables. The 1-norm is converted as follows:

$$\begin{cases}\kappa_s^++\kappa_s^-,,1\\ \sum_{s=1}^{N_s}\left(p_s^++p_s^-\right)\leq\theta_1\\ 0,,p_s^+,,\kappa_s^+\theta_1\\ 0,,p_s^-,,\kappa_s^-\theta_1\end{cases}\quad\forall s \tag{48}$$

where $p_s^+$ and $p_s^-$ are the positive and negative offsets of the probability distribution $p_s$, respectively, compared with the original probability distribution; and $\kappa_s^+$ and $\kappa_s^-$ are 0-1 flags for positive and negative offsets, respectively.

Similarly, the ∞-norm can be linearized with Boolean variables $\gamma_s^+$ and $\gamma_s^-$:

$$\begin{cases} \gamma_s^+ + \gamma_s^- \leq 1 \\ p_s^+ + p_s^- \leq \theta_\infty \\ 0 \leq p_s^+ \leq \gamma_s^+ \theta_\infty \\ 0 \leq p_s^- \leq \gamma_s^- \theta_\infty \end{cases} \forall s \quad (49)$$

$$p_s = p_{s0} + p_s^+ + p_s^- \quad (50)$$

Finally, $p_s$ is the actual scenario probability distribution.

## 5 Model solution

### 5.1 WGAN-GP model training process

The training process of the WGAN-GP is as follows:

Step 1: Random noise $z$ is fed into the generator. According to the distribution of historical samples $x$, the generator is trained to generate random samples;

Step 2: The generated samples and historical samples $x$ are sent to the discriminator simultaneously, and then the discriminator outputs the probability that the generated samples are real samples;

Step 3: After calculating the loss functions of the generator and discriminator, the weights of the generator and discriminator networks are respectively updated;

Step 4: The generator and discriminator are iteratively optimized until the end of training.

### 5.2 DRO-model solving steps

To solve the proposed DRO model, the column and constraint generation (CCG) algorithm has the advantages of few iterations and high accuracy [37].

Eq. (39) must be decomposed into the main problem (master problem, MP) and sub-problem (sub-problem, SP) and iterations must be completed until the targeted iteration accuracy is satisfied. The MP is expressed as follows:

$$\text{MP:} \quad \min_{u \in U, p_s \in \Omega^p, v_s \in V} Au + \eta \quad (51)$$

$$\eta \geq \sum_{s=1}^{N_s} p_s^w B v_s^w, \quad w=1,2,\ldots,W \quad (52)$$

where $w$ represents the number of iterations, W denotes the maximum number of iterations, and $\eta$ is the introduced auxiliary variable.

The MP finds the optimal solution of the first-stage objective function according to the known worst probability distribution value, and the lower bound value of the model is updated. The SP is expressed as follows:

$$\text{SP:} \quad F(u^*) = \max_{p_s \in \Omega^p} \sum_{s=1}^{N_s} p_s^w \min_{v_s \in V} B v_s^w, \quad w=1,2,\ldots,W \quad (53)$$

The SP is solved under the variable $u_w^*$ given by the MP to obtain the worst probability distribution of the current iteration, which is fed back to the MP for the next iteration, and the upper bound value is updated for Eq. (52).

When solving the SP specifically, the objective function is expressed as a max-min bi-level optimization problem. However, given that there is no direct coupling relationship between the probability value of RGs in the SP and the second-stage decision variables, the SP can be solved in two steps. The MP is solved first, and then the SP is solved.

The model solving process is shown in Fig. 4.

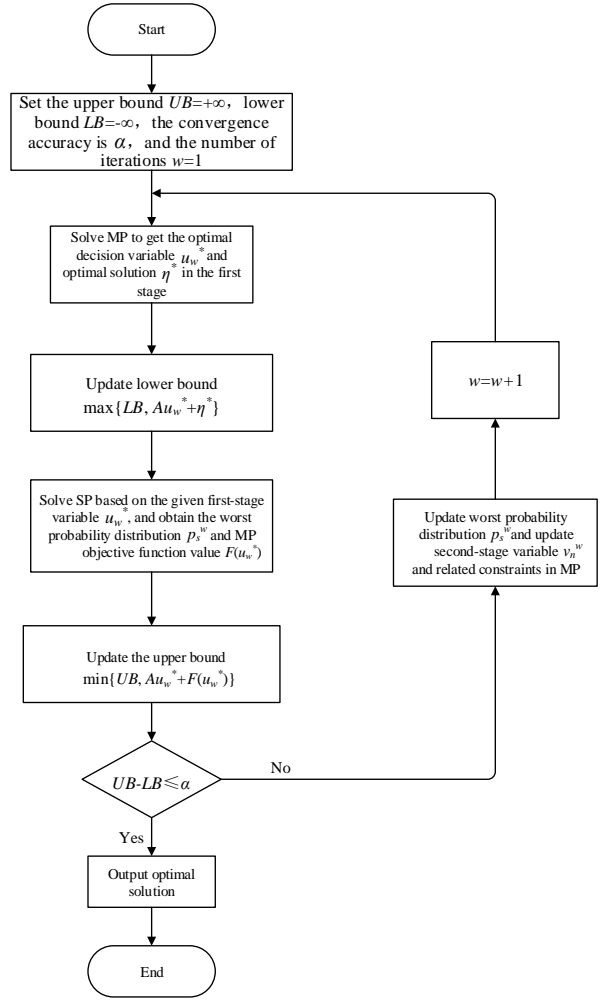

**Fig. 4. CCG solving process**

Step 1: Set the upper bound $UB=+\infty$, lower bound $LB=-\infty$, convergence accuracy as $\alpha$, number of iterations $w=1$, and initial probability distribution of the second stage $p_{s0}$;

Step 2: Solve the MP. Obtain the optimal decision variable in the first stage; the optimal solution simultaneously updates the lower bound $\max\left\{LB, Au_w^* + \eta^*\right\}$;

Step 3: Solve the SP on the basis of the given first-stage variable $u_w^*$, obtain the worst probability distribution $p_s^w$ and SP objective function value $F\left(u_w^*\right)$, and update the upper bound $\min\left\{UB, Au_w^* + F\left(u_w^*\right)\right\}$ simultaneously;

Step 4: If $UB - LB \leq \alpha$, terminate the iterative process and output the optimal solution; otherwise, update the worst probability value $p_s^w$ and update the second-stage decision variable $v_n^w$ and related constraints in the MP;

Step 5: Update $w=w+1$, and return to Step 2 to continue the iteration.

## 6 Case study
### 6.1 Simulation system and data

To test the effectiveness of the proposed DRO model and solution algorithm, an actual CIES in North China was selected for simulation analysis, and the optimization program was solved by using MATLAB R2016b coupled with IBM ILOG CPLEX Optimizer. The 2-year measured data of PV and WT outputs in a certain area in North China were used as the historical data set; the sampling interval was 15 min. The data set was split into 80% for training and 20% for testing. In the WGAN-GP, the learning rate was set to 0.0002 [38]. Table 1 lists the structure and parameters of the generator and discriminator. Table 2 shows the time-of-use (TOU) price of the power grid in this region, and Table 3 indicates the operating parameters of the CIES equipment.

**Table 1. Structure and parameters of generator and discriminator**

| Type | layer | Title | Parameters | Numerical |
|---|---|---|---|---|
| Generator | 1 | Fully connected layer | Number of neurons | 128 |
| | | Activation function | ReLU | — |
| | 2 | Fully connected layer | Number of neurons | 256 |
| | | Batch norm | Dynamic mean momentum | 0.8 |
| | | Activation function | ReLU | — |
| | 3 | Fully connected layer | Number of neurons | 512 |
| | | Batch norm | Dynamic mean momentum | 0.8 |
| | | Activation function | ReLU | — |
| | 4 | Fully connected layer | Number of neurons | 1024 |
| | | Batch norm | Dynamic mean momentum | 0.8 |
| | | Activation function | ReLU | — |
| | 5 | Fully connected layer | Number of neurons | 24*24*1 |
| | | Activation function | tanh | — |
| Discriminator | 1 | Fully connected layer | Number of neurons | 512 |
| | | Activation function | LeakyReLU | 0.2 |
| | 2 | Fully connected layer | Number of neurons | 256 |
| | | Activation function | LeakyReLU | 0.2 |
| | 3 | Fully connected layer | Number of neurons | 1 |

**Table 2. Grid TOU price**

| Periods | Specific time periods | | Electricity prices (¥/kWh) |
|---|---|---|---|
| Peak period | 9:00-11:00 | 19:00-23:00 | 1.35 |
| Flat period | 8:00 | 12:00-18:00 | 0.90 |
| Valley period | 1:00-7:00 | 24:00 | 0.48 |

**Table 3. CIES operating parameters**

| Parameter | Value | Parameter | Value |
|---|---|---|---|
| $P_{MTG,EL}^{\max}$ (kW) | 300 | $C_0^{ESS}/C_{\min}^{ESS}$ (kWh) | 10 |
| $P_{U,\max}^{MTG}$ (kW) | 50 | $\eta_{ch}/\eta_{dc}$ | 0.95/0.95 |
| $P_{D,\max}^{MTG}$ (kW) | 50 | $C_{\max}^{ESS}$ (kWh) | 90 |
| $P_{MTG,HL}^{\max}$ (kW) | 360 | $P_{HL}^{EB}$ (kW) | 200 |
| $\eta_{HE}$ | 1.2 | $\eta_{EB}$ | 0.9 |
| $P_{CH,\max}^{ESS}/P_{DC,\max}^{ESS}$ (kW) | 20/20 | $P_{\min}^{grid}$ (kW) | 0 |
| $a_i$ | 1.2 | $P_{\max}^{grid}$ (kW) | 600 |
| $b_i$ (¥/kW) | 0.0015 | $c_{ESS}$ (¥/kW) | 0.02 |
| $S_{T_i}$ (¥/kW) | 0.25 | $c_{Loss}$ (¥/kW) | 0.62 |
| $S_{K_i}$ (¥/kW) | 0.25 | $c_{HSD}$ (¥/kW) | 0.011 |
| $\eta_{ch}^{HSD}/\eta_{dc}^{HSD}$ | 0.85/0.9 | $P_{CH,\max}^{HSD}/P_{DC,\max}^{HSD}$ (kW) | 50/50 |
| $k_{MTG}$ (kg/kW) | 0.49 | $k_{grid}$ (kg/kW) | 0.82 |

## 6.2 Performance Evaluation of the WGAN-GP

(1) Wasserstein distance: the goal of model training is to make the loss as small as possible, that is, to minimize the Wasserstein distance between the true and generated distributions [39]. This is expressed as follows:

$$\text{WD}(p_d, p_g) = \min_{w \in \mathbb{R}^{n \times m}} \sum_{i=1}^{n} \sum_{j=1}^{m} w_{ij} d(x_i^d, x_j^g) \quad (54)$$

$$\text{s.t.} \sum_{j=1}^{m} w_{i,j} = p_d(x_i^d) \ \forall i, \ \sum_{i=1}^{n} w_{i,j} = p_g(x_j^g) \ \forall j \quad (55)$$

where $d(x_i^d, x_j^g)$ is the reference distance between the real data and generated samples. In the initial stage of training, the generated samples significantly differ from the real data. The generator adjusts the weights to make both sets more similar, and the discriminator also improves the discrimination ability through training. As the Wasserstein distance continues to decrease, eventually the discriminator will not be able to accurately distinguish the source of the input samples, and the RG output scenario will fully reflect the true distribution of historical data.

(2) Fréchet inception distance (FID) measures the distance between real and synthetic samples [40]:

$$FID(P_x, P_g) = \|\mu_x - \mu_g\| + Tr\left[\sigma_x^2 + \sigma_g^2 - 2(\sigma_x^2 \sigma_g^2)^{\frac{1}{2}}\right] \quad (56)$$

where $\mu_x$ and $\mu_g$ are the mean values of the real data and generated sample distributions, respectively; and $\sigma_x^2$ and $\sigma_g^2$ are the covariances of the real data and generated sample distributions, respectively. The smaller the FID, the closer both distributions are, which means better training performance.

(3) Maximum mean discrepancy (MMD) is a measure of the difference between two distributions in a Hilbert space. It can be employed for measuring the distance between the generated and real data sets. MMD represents the difference between $P_d$ and $P_g$ for some fixed kernel function $k$, which is defined as follows [20]:

$$MMD^2(P_d, P_g) = E_{x_d, x_d' \sim P_d, x_g, x_g' \sim P_g}\left[k(x_d, x_d') - 2k(x_d, x_g) + k(x_g, x_g')\right] \quad (57)$$

Similar to FID, the smaller the MMD distance, the better the generation performance, which means the better the performance of the WGAN-GP.

The changes of these three evaluation metrics during the training process are shown in Fig. 5.

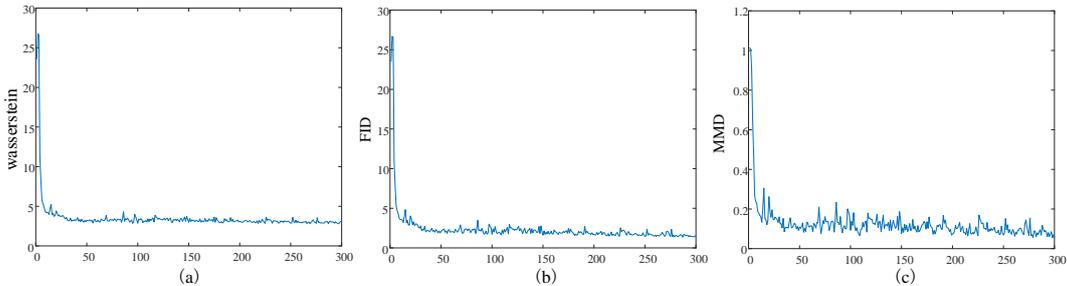

**Fig. 5. Changes of the evaluation metrics during training**

## 6.3 Clustering Results

DBI is a minimum value index, that is, the clustering algorithm has the best classification effect when the DBI reaches its minimum value. The value range of the silhouette coefficient is [-1, 1]; the closer it is to 1, the higher the intra-cluster similarity, the lower the inter-cluster similarity, and the better the clustering effect. It can be seen from Fig. 6 that for the PV output, the optimal number of clusters is 2; for the WT output, the optimal number of clusters is 4. Fig. 7 shows typical RG output scenarios obtained through scenario generation and clustering.

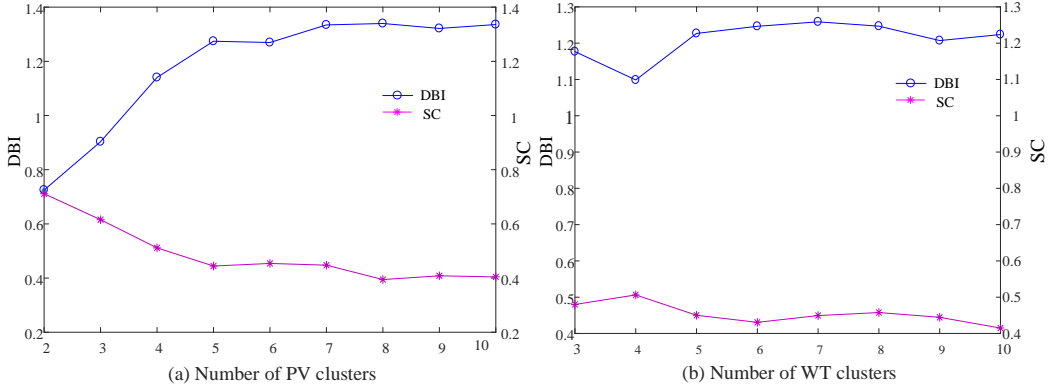

**Fig. 6. DBI and SC values for different number of clusters**

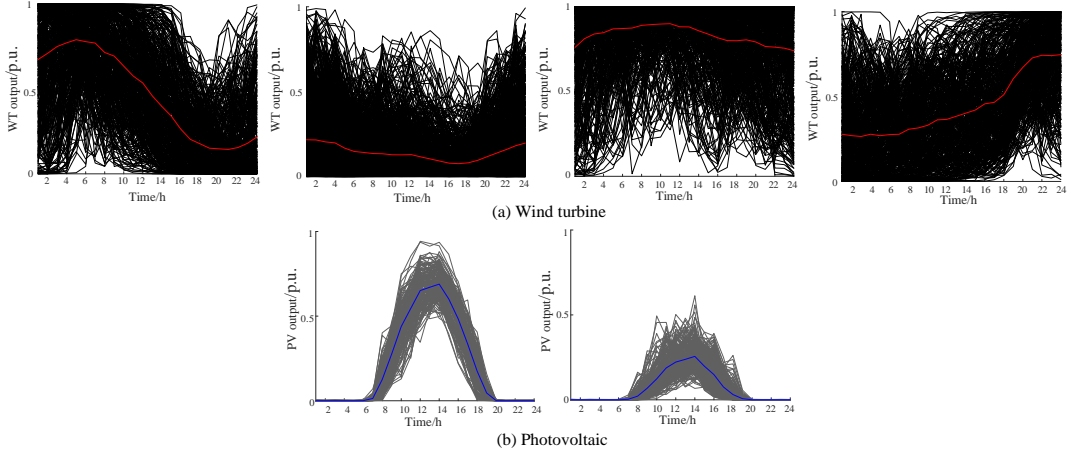

**Fig. 7. Optimal clustering results of RG outputs**

### 6.4 Analysis of Scheduling Results in Typical RG Scenarios

To verify the effectiveness of the proposed DRO model, the scheduling scheme of typical RGs scenarios was analyzed, setting $M$=5000, $\alpha_1$=0.99, and $\alpha_\infty$=0.99. The results of CIES electrical and heat scheduling in typical scenarios are shown in Figs. 8 and 9, respectively.

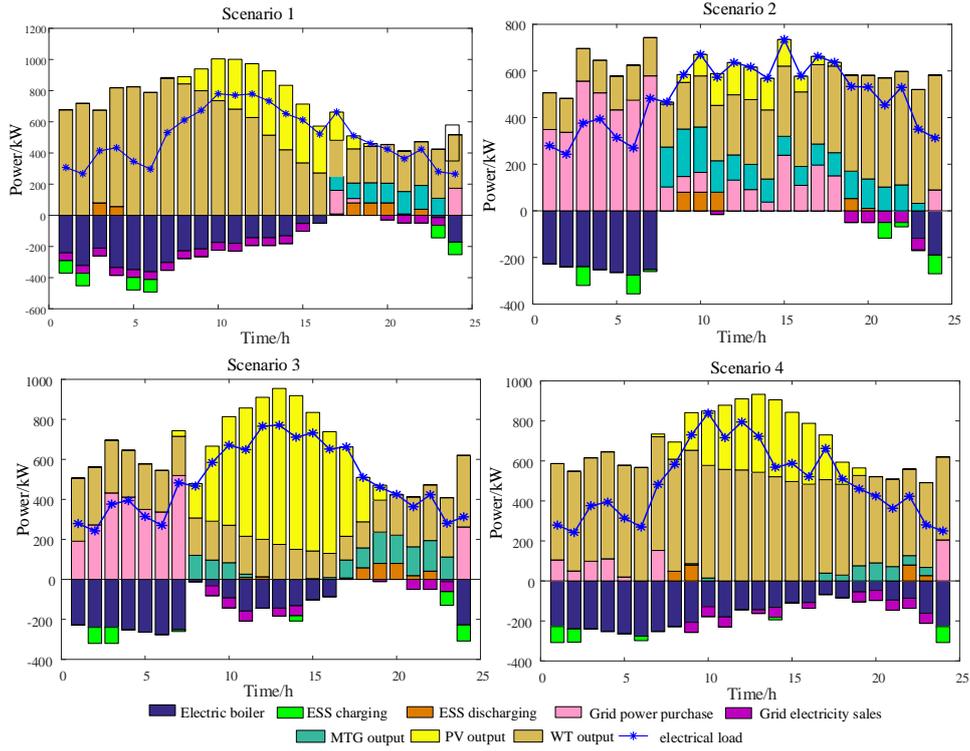

**Fig. 8. Results of CIES electrical scheduling in typical scenarios**

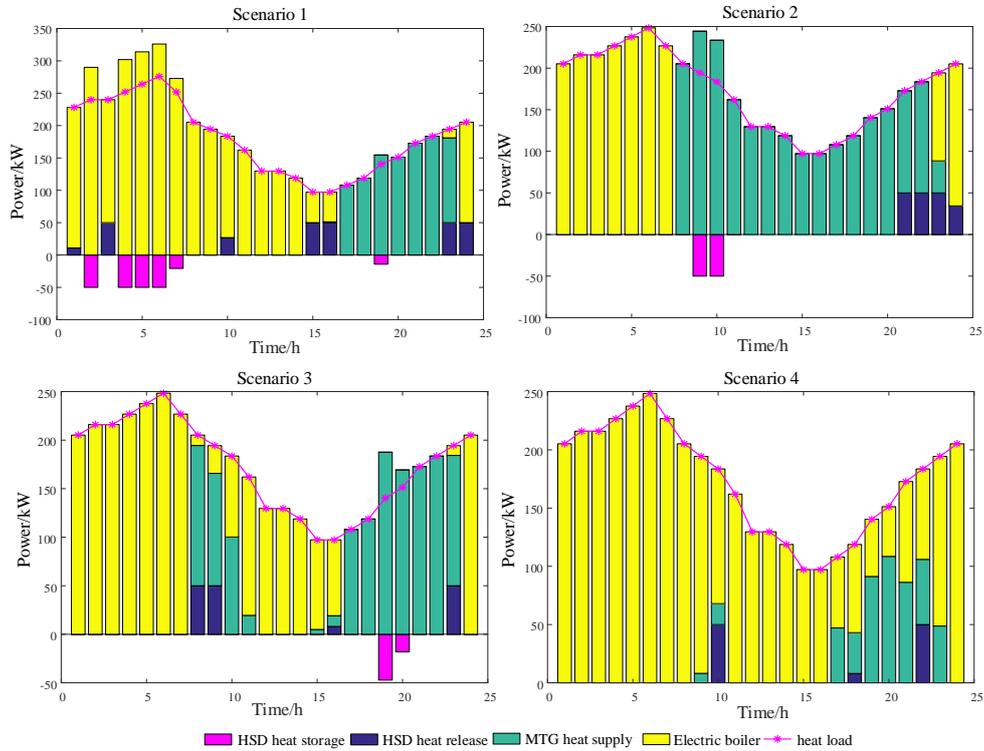

**Fig. 9. Results of CIES heat scheduling in typical scenarios**

It can be seen from Fig. 8 that, in terms of electric energy, in the periods when the RG outputs were sufficient, the electric load was entirely supplied by renewable energy (e.g., the 9:00-16:00 periods in Scenarios 1 and 4). During peak load periods, such as 8:00-10:00 and 18:00-21:00, the ESS gave priority to release electric energy, but its capacity was limited, so the power vacancy also needed to be satisfied by the MTG and grid. According to the scheduling results, the power generation and purchase costs of the CIES were reduced on the basis of prioritizing RGs consumption, and the bidirectional energy flow of the ESS also reduced the power supply pressure, which in turn decreased the electricity curtailment rate, thereby effectively achieving a more economical system operation.

Fig. 9 shows that, in terms of heat energy, the total heat load was satisfied by the MTG and EB. During most of the daytime, the MTG was put into operation to compensate for the power gap of RG. The heat load was mainly borne by the MTG power-supply waste heat in co-generation mode. During the night time, the heat load was basically supplied by the EB. Given that the electric load was in the valley period but the heat load was at the peak period, the heat supply of MTG was insufficient, and the heat load could only be borne by the EB. In addition, HSD, as a buffer device, also plays an important role in the balance of heat supply and demand.

**6.5 Integrated Demand Response Analysis**

To analyze the results of electrical and heat loads before and after integrated demand response. Fig. 10 shows the load curves before and after integrated demand response in typical scenarios.

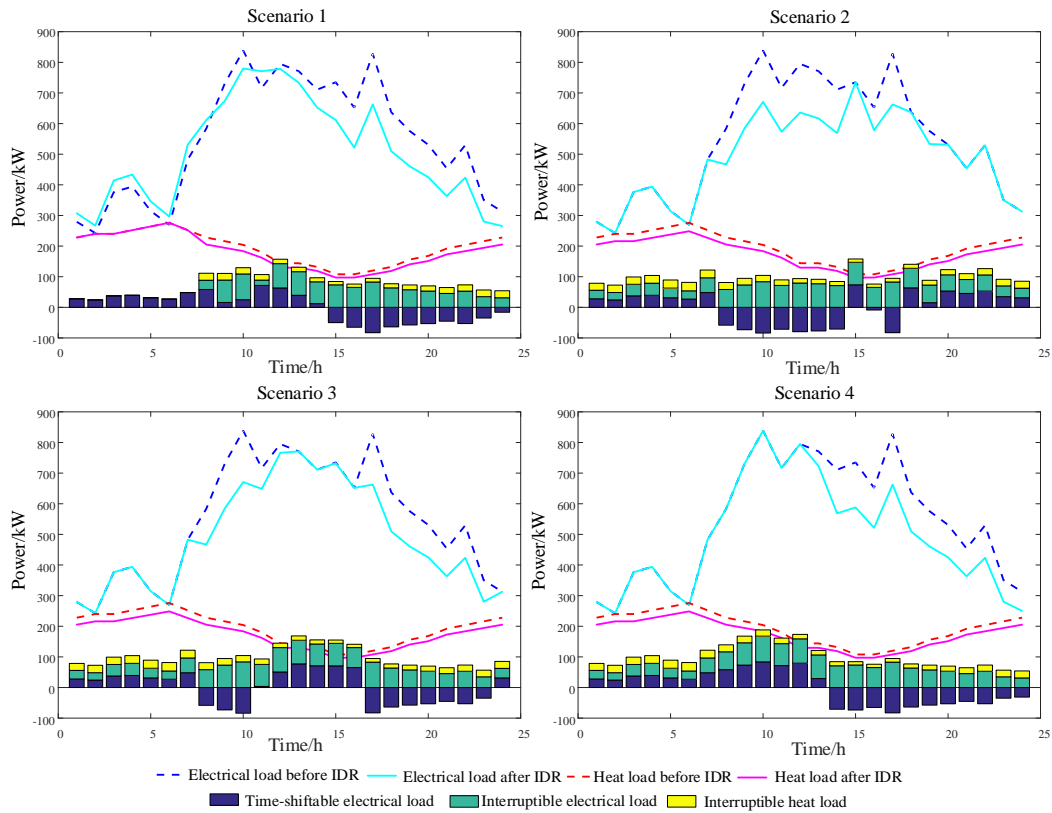

**Fig. 10. Results of IDR scheduling in typical RG output scenarios**

It can be seen from Fig. 10 that after the application of integrated demand response, the electric load curve tended to be smooth, realizing peak shaving and valley filling. During peak electricity consumption periods (8:00–12:00 and 18:00–21:00), under the influence of incentive policies and electricity prices, users tended to decrease electricity consumption by shifting and interrupting electricity loads. During low electricity load periods (1:00-6:00), the lower electricity price led to an increase in electricity consumption. In addition, during the periods 20:00–24:00 and 1:00-5:00, the heat-load reduction phenomenon was noticeable. This is because the users are less sensitive to temperature when sleeping at night, and the comfort requirements can be appropriately relaxed.

**6.6 Economic Analysis**

To verify the validity and superiority of the proposed model, four operating modes are set up for comparison. In Modes 1 and 2, IDR is not applied, and the heat load is set to the value at PMV=0. Mode 3 represents the DRO model proposed in this paper. Table 4 indicates the optimization results under the four modes.

Mode 1: Multiple RGs uncertainties are considered, and IDR is not applied;

Mode 2: Multiple RGs uncertainties are not considered, and IDR is not applied;

Mode 3: Multiple RGs uncertainties are considered, and IDR is applied;

Mode 4: Multiple RGs uncertainties are not considered, and IDR is applied.

Table 4. Comparison of results for different operating modes

| Optimization results | Operating modes | | | |
| --- | --- | --- | --- | --- |
| | Mode 1 | Mode 2 | Mode 3 | Mode 4 |
| MTG operation cost /¥ | 1883.23 | 1795.63 | 1730.26 | 1649.23 |
| Grid power purchase cost /¥ | 973.41 | 922.25 | 860.85 | 678.36 |
| Grid power sales profit /¥ | 245.21 | 190.63 | 196.66 | 201.79 |
| ESS operating cost/¥ | 24.23 | 28.65 | 30.36 | 25.15 |
| HSD operating cost/¥ | 27.25 | 24.36 | 25.26 | 30.71 |
| Carbon penalty cost/¥ | 186.56 | 177.56 | 170.45 | 149.52 |
| IDR cost /¥ | — | — | 154.36 | 274.79 |
| Renewable power curtailment cost /¥ | 191.35 | 123.32 | 142.86 | 92.08 |
| Total operating cost /¥ | 3039.65 | 2879.36 | 2914.56 | 2696.78 |
| Renewable power curtailment rate /% | 14.81 | 10.01 | 12.07 | 8.16 |

From Table 4, it can be seen that compared with the deterministic optimization, the DRO produces a scheduling scheme with higher total operating costs, but lower renewable power curtailment rates. (1) Due to the consideration of multiple RG uncertainties, the DRO scheduling schemes require a higher operating cost to deal with possible worst scenarios, but doing so effectively reduces the impact of randomness and the volatility of RG outputs, and improves the accuracy of the scheduling schemes; while the deterministic optimization makes decisions based on RG predicted outputs, and the resulting scheduling schemes are risky and lack the adaptability to the uncertainty of system operation. (2) The renewable power curtailment rate decreases from 14.81% under Mode 1 to 10.01% under Mode 2. This is because the deterministic method is only used for scheduling specific RG output forecasts, and the forecast uncertainty is ignored, which is quite different from the actual situation, thus leading to a higher power procurement rate. The similar conclusion can be reached when comparing Modes 3 and 4. Therefore, considering the uncertainty of RGs could improve the utilization of RGs to a certain extent.

Table 4 also suggests the IDR mechanism can improve the economy of system operation. Specifically, the operating costs are reduced from 3039.65¥ under Mode 1 to 2914.56¥ under Mode 3, and from 2879.36¥ under Mode 2 to 2696.78¥ under Mode 4. Owing to the peak-shaving and valley-filling effect of the flexible loads, the electricity price constitutes an incentive signal to guide users to appropriately adjust electricity and heat consumption. By this means, the total operating cost is reduced by implementing IDR programs.

**6.7 Sensitivity analysis of ambiguity set parameters**

**6.7.1 Analysis of the number of historical data**

To analyze the impact of the number of historical data on the CIES operating cost, we set $\alpha_1 = 0.5$, $\alpha_\infty = 0.99$. The CIES operating costs under different historical data numbers are shown in Table 5.

**Table 5. Comparison of operating costs under different numbers of historical data**

| Number of data | operating costs/¥ | | |
| --- | --- | --- | --- |
| | 1-norm | ∞-norm | Comprehensive norm |
| 100 | 2969.75 | 3185.45 | 2945.34 |
| 1000 | 2715.10 | 2736.67 | 2762.53 |
| 2000 | 2700.95 | 2711.74 | 2712.66 |
| 5000 | 2694.47 | 2699.78 | 2691.98 |
| 10000 | 2689.64 | 2691.79 | 2689.39 |
| 20000 | 2688.22 | 2689.30 | 2688.10 |

Table 5 shows that as the number of historical data increases, the total operating cost of the CIES gradually decreases. When the number of historical data reaches a certain level, the decreasing amplitude of the total operating cost becomes smaller. This is because the more historical data there are, the closer the probability distribution of the initial RG scenario is to the real output. As a result, the shiftable range of the worst-case scenario probability is reduced, requiring less scheduling cost to manage worst-case scenarios.

**6.7.2 Analysis of different confidence levels**

To analyze the effect of confidence levels on the CIES operating costs, we set $M=5000$. The operating costs of the CIES under different confidence levels are shown in Table 6.

**Table 6. Operating costs for different confidence levels**

| $\alpha_\infty$ | Operating costs /¥ | | |
| --- | --- | --- | --- |
| | $\alpha_1=0.2$ | $\alpha_1=0.5$ | $\alpha_1=0.99$ |
| 0.5 | 2686.54 | 2688.94 | 2690.94 |
| 0.8 | 2690.84 | 2690.94 | 2692.31 |
| 0.99 | 2691.23 | 2691.98 | 2696.78 |

Note that the CIES operating costs keep rising as confidence levels $\alpha_1$ and $\alpha_\infty$ increase. The reason is that the higher the confidence level, the larger the confidence interval. This leads to increased uncertainty and higher robustness of the operating scheme.

To further analyze the superiority of the comprehensive-norm confidence interval, the optimization results obtained by considering only the 1-norm or ∞-norm constraints are compared with those obtained by considering the comprehensive-norm constraints. The specific parameters were set as follows:

(1) A comparison of the operating costs with different values of the comprehensive norm and only ∞-norm (with $\alpha_\infty = 0.99$) are shown in Table 7.

**Table 7. Operating costs for different values of the comprehensive norm and ∞-norm**

| $\alpha_1$ | Operating costs /¥ | |
| --- | --- | --- |
| | comprehensive norm | ∞-norm |
| 0.2 | 2691.23 | 2699.78 |
| 0.5 | 2691.98 | 2699.78 |
| 0.99 | 2696.78 | 2699.78 |

Table 7 indicates that when the number of historical data remains unchanged, the optimization result under the confidence interval of the comprehensive norm is more economical than when only the ∞-norm is considered. This is because the comprehensive-norm confidence interval is smaller and the RGs uncertainty can be more accurately characterized, resulting in lower operating costs.

(2) A comparison of the operating costs with different values of the comprehensive norm and only 1-norm (with $\alpha_1 = 0.5$) is shown in Table 8.

Similarly, it can be seen from Table 8 that when the number of historical data remains unchanged, the optimization result under the confidence interval of the comprehensive norm is less conservative than when only the ∞-norm is considered. This means that the comprehensive norm as a confidence interval is closer to real RG output scenarios.

Table 8. Operating costs for different values of the comprehensive norm and ∞-norm

| $\alpha_\infty$ | Operating costs/¥ | |
|---|---|---|
| | Comprehensive norm | 1- norm |
| 0.5 | 2688.94 | 2694.47 |
| 0.8 | 2690.94 | 2694.47 |
| 0.99 | 2691.98 | 2694.47 |

**6.8 Computational complexity analysis**

To verify the computational complexity of the CCG algorithm, the number of historical data is set to 5000, and $\alpha_1$ and $\alpha_\infty$ are respectively selected as 0.5 and 0.99. The change of results during the iteration of the CCG algorithm is shown in Fig. 11.

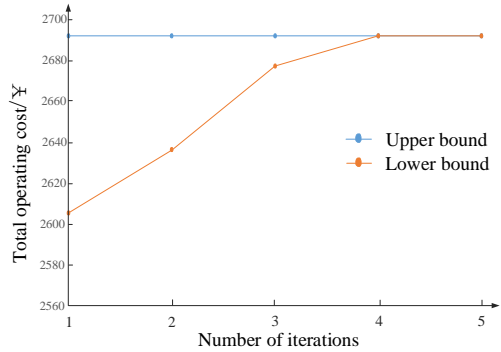

Fig. 11 The change of results during iterations

It can be seen from Fig. 11 that the objective function value has met the given iteration accuracy after four iterations. This is because the DRO model is decoupled into MP and SP using the CCG algorithm, where there is no direct coupling between the scenario probability distribution in SP and the second stage decision variables. So, it is unnecessary to use the traditional strong duality or Karush-Kuhn-Tucker (KKT) conditions to convert the bi-level model of the SP into a single-level model, which avoids the complex model transformation problem. This reduces the computational complexity and can meet the real-time requirements of actual scheduling.

**6.9 Comparison with other optimization methods**

**6.9.1 Comparison with stochastic programming and robust optimization**

In this section, the proposed DRO method is compared with traditional stochastic programming and robust optimization methods. The robust optimization approach uses a box uncertainty set and sets the fluctuation interval as 20%; stochastic programming adopts typical scenarios and corresponding probability distributions. The DRO model parameters were set as follows: $M$=5000, $\alpha_1 = \alpha_\infty = 0.99$. The optimization results of these methods are shown in Table 9.

**Table 9. Optimization results of different optimization methods**

| Costs/Profits | Optimization methods | | |
| --- | --- | --- | --- |
| | Stochastic programming | Robust optimization | Proposed DRO |
| MTG operation cost /¥ | 1523.23 | 1735.63 | 1649.23 |
| Grid power purchase cost /¥ | 658.36 | 689.25 | 678.36 |
| Grid power sales profit /¥ | 222.55 | 211.12 | 201.79 |
| ESS operating cost/¥ | 28.69 | 30.53 | 25.15 |
| HSD operating cost/¥ | 32.25 | 25.66 | 30.71 |
| Carbon penalty cost/¥ | 136.45 | 168.55 | 149.52 |
| IDR cost /¥ | 276.89 | 280.36 | 274.79 |
| Renewable power curtailment cost /¥ | 108.53 | 150.34 | 92.08 |
| Total operating cost /¥ | 2539.65 | 2866.42 | 2696.78 |
| Renewable power curtailment rate /% | 10.86 | 13.66 | 8.16 |

Note that the total operating cost of the CIES optimized by the DRO model is lower than that of traditional robust optimization and higher than that of stochastic programming. This is because the scenarios obtained by stochastic programming based on the probability distribution function do not consider the worst cases, resulting in weak robustness; traditional robust optimization pays too much attention to extreme RGs scenarios, which causes the CIES scheduling scheme to be too conservative, resulting in higher curtailment cost and less economical operation than the DRO method.

All in all, DRO combines the characteristics of stochastic programming and robust optimization, making full use of the scenario probability distribution generated by a large amount of historical data. The RGs uncertainty can be managed by adjusting the dispatching scheme including the MTG output, grid power purchase, ESS charge and discharge, etc. Compared with robust optimization considering only worst scenario information, the DRO model improves the consumption rate of renewable energy and achieves a more economical operation. Therefore, DRO can effectively avoid the limitations of stochastic programming and robust optimization, and achieve a good balance between economy and robustness.

**6.9.2 Comparison with other DRO methods**

In order to further test the performance of the proposed DRO method, comparative tests have also been implemented between ours and other existing DRO methods. Considering that Moment based DRO (MDRO) and Wasserstein based DRO (WDRO) are commonly used DRO methods to deal with the uncertainty of renewable energy [13], which are adopted as the comparison algorithms, where the moment based ambiguity set contains all distributions with the prescribed moment information (mean and covariance). The optimization results of different DRO methods are listed in Table 10.

**Table 10. Optimization results of different DRO methods**

| DRO methods | Total operating cost /¥ | Renewable power curtailment rate /% | Calculation time /s |
|---|---|---|---|
| Proposed DRO | 2696.78 | 8.16 | 16.8 |
| MDRO | 2795.23 | 9.68 | 286.2 |
| WDRO | 2706.92 | 8.55 | 370.7 |

As shown in Table 10, the performance of the proposed DRO method outperforms the other DRO methods, which is reflected by the fact that our method has less operating cost, lower renewable power curtailment rate, and less calculation time. (1) When enough renewable power output data are available, more probability information can be obtained, but only moment information is included in the moment based fuzzy set, which cannot reduce the ambiguity set by merging more data. Thus, the MDRO is more conservative than ours and the WDRO. (2) Regarding the total operating cost, the proposed DRO has little difference with the WDRO in conservatism, but has significant advantages over the WDRO and MDRO in calculation time. The reason for this is that the WDRO and MDRO generally construct convex optimization model through the reflected linear decision rule and the duality theorem, which will increase the difficulty of optimization model transformation and its solution; while our method only needs a simple linearization, which avoids a complex mathematical conversion and calculation process. Therefore, it can be concluded that our method is superior to other commonly used DRO methods.

## 7 Conclusions

To manage the uncertainty of renewable energy generation and promote the consumption of renewable energy, a data-driven two-stage distributed robust scheduling model for CIESs was developed. A comprehensive norm consisting of the 1-norm and ∞-norm was defined as a probability distribution confidence set. In the solving stage of the proposed model, the CCG algorithm was adop1ted to iteratively solve the DRO model, thereby avoiding the complex transformation of strong duality or KKT conditions and reducing the complexity of the solution. The main conclusions drawn are as follows:

(1) To manage multiple RGs uncertainties, a scenario generation method based on a WGAN-GP is proposed. This solves the gradient vanishing and mode collapse problems of original GANs and gives rise to a more stable training. The historical data set of RG outputs is fully utilized. This avoids traditional probability density information, which leads to a wider applicability.

(2) The PMV metric is introduced to describe the ambiguity of users' thermal comfort. An integrated demand response mechanism was devised to realize peak shaving and valley filling and reduce CIES operating costs.

(3) By analyzing the sensitivity of the ambiguity set parameters, we observed that the larger the historical data set of RGs, the smaller the confidence interval of the uncertainty set, and the lower the conservativeness of the scheduling plan. Compared with only considering the ∞-norm or 1-norm, the comprehensive norm employed as the confidence interval is closer to real RG output scenarios, and the resulting operation is more economical.

(4) With respect to traditional stochastic and robust methods, the simulations performed on an actual CIES in North China indicate that the proposed DRO model can well balance the relationship between system economy and robustness, thereby confirming the effectiveness and superiority of the DRO model. Furthermore, our approach is better than other commonly used DRO methods, i.e. Moment based DRO and Wasserstein based DRO, with better operational economy, lower renewable power curtailment rate, and higher computational efficiency.

The presented work has not taken into account the data privacy protection issue [41], while more realistic application scenarios should consider the preservation of data privacy while designing CIES scheduling schemes. With the increasing prevalence of cyber-attacks, e.g., false data injection attacks [42], it would be interesting in the future to find out how to make the scheduling strategies resilient to these attacks.

**Acknowledgment**

This work is supported by the National Natural Science Foundation of China under Grant No. U2066208, and the Natural Science Foundation of Jilin Province, China under Grant No. YDZJ202101ZYTS149.